\documentstyle[12pt]{article}
\begin{document}

\newtheorem{theorem}{Theorem}
\newtheorem{lemma}[theorem]{Lemma}
\newtheorem{corollary}[theorem]{Corollary}
\newtheorem{definition}[theorem]{Definition}
\newtheorem{remark}[theorem]{Remark}

\def\ov{\overline}
\def\var{\varepsilon}
\def\om{\omega}
\def\pa{\partial}
\def\la{\langle}
\def\ra{\rangle}
\def\om{\omega}

\title{On the structure  of Markov flows}
\author{L. Accardi, S.V. Kozyrev}
\maketitle

\centerline{\it Centro Vito Volterra Universita di Roma Tor Vergata}

\begin{abstract}
A new infinitesimal characterization of completely positive
but not necessarily homomorphic Markov flows from a $C^*$--algebra
to bounded operators on the boson Fock space over $L^2(  R)$ is given.
Contrarily to previous characterizations, based on stochastic
differential equations, this characterization is universal,
i.e. valid for arbitrary Markov flows.
With this result the study of Markov flows
is reduced to the study of four $C_0$--semigroups.
This includes the classical case and even in this case
it seems to be new.
The result is applied to deduce a new existence theorem for Markov
flows.
\end{abstract}

\section{Introduction}

Our goal is to understand the structure of dynamical evolutions, both
classical and quantum. In both cases a reversible dynamical evolution
is described by a $1$--parameter group of automorphisms of a certain
algebra and an irreversible one by a $1$--parameter semi--group of
completely positive maps on the same algebra.

Perturbations of reversible free evolutions lead to introduce the so
called {\it interaction picture}. Mathematically this leads to
generalize the notion of $1$--parameter group into that of {\it flow}.
Flows with an additional covariance property are $1$--cocycles for the
free evolution. Flows with an additional localization property are
called {\it Markov flows} or {\it Markov cocycles} and were introduced
in \cite{Ac78}.
Summing up: the problem of studying the structure of nonlinear dynamical
evolutions is equivalent to determine the structure of flows.
Smooth deterministic equations lead to usual flows; stochastic or white
noise equations lead to Markov flows.

A particular class of Markov flows are the Markov semigroups. Their
structure, in the strongly continuous case, is determined, in the abstract
context of Banach spaces, by the Hille--Yoshida theorem which characterizes
their generators in terms of dissipations (derivations in the reversible
case). In the present paper we obtain a characterization of strongly
continuous Markov flows in terms of infinitesimal characteristics.
We solve this problem in the context of Markov flows from a
$C^*$--algebra ${\cal B}_S$ to the algebra
${\cal B}_S\otimes {\cal B}(\Gamma (L^2(  R)))$
of bounded operators on the boson Fock space over $L^2(  R)$,
identified to the standard Wiener space.

% White noise equations are introduced as a natural way to describe in a
% unified manner these four semigroups.

Since it is known that any stochastic process satisfying a
(classical or quantum) stochastic differential equation admitting an
existence, uniqueness and regularity theorem for a sufficiently
large class of initial data, gives rise to a Markov flow,
a corollary of our result is an infinitesimal characterization
of such processes.

The standard way to attack this problem up to now has been to
show that a flow satisfies a stochastic differential equation, of the
type first considered by Evans and Hudson \cite{EvHu88},
and to consider the structure maps defining such an equation as
the infinitesimal characteristics of a flow.

Our approach is different: to every strongly continuous flow
on a $C^*$--algebra ${\cal B}_S$ we associate a
completely positive (but not identity preserving)
$C_0$--semigroup on the $2\times 2$ matrices with
coefficients in ${\cal B}_S$. This gives four $C_0$--semigroups on
${\cal B}_S$: the infinitesimal characteristics of the flow
are the generators of these semigroups.

If the flow satisfies a stochastic differential equation, of
Evans--Hudson type, then it is easy to express their structure maps in
terms of our generators and conversely. However our generators always
exist, while the existence of the structure maps is constrained by
strong analytical conditions that are neither easy nor natural to
formulate in terms of the flow itself.

The structure of the present paper is the following. In section (2)
we remind some general properties of Markov flows.
Starting from section (3) we specialize our context to flows on
the boson Fock space $\Gamma(L^2(R,{\cal K}))$
and we prove that such a flow is uniquely determined by a family
$(P^{s,t}_{f,g})$ of $C_0$--evolutions, indexed by pairs of
elements $f,g$ in a totalizing set of $L^2(R,{\cal K})$
(cf. Definition (3) and Theorem (12) below). In section (4)
we show that the evolutions $(P^{s,t}_{f,g})$ reduce to semigroups
$(P^{t}_{f,g})$ in the case of covariant flows.
In section (5), using a known result on exponential vectors,
we show that the family $(P^{t}_{f,g})$ of semigroups
can in fact be reduced to four semigroups which allows to define
a single completely positive $C_0$--semigroup on the algebra
$M(2,{\cal B}_S)$ of $2\times 2$ matrices with coefficients in
${\cal B}_S$. In section (6) we further specialize to the class
of flows satisfying a stochastic differential equation
and prove a general existence theorem applicable to infinite lattice
spin systems, whose analysis \cite{AcKo99a} motivated the present paper.

\section{Markov flows}

In this section we recall some general properties of Markov flows.
For more information we refer to \cite{AcMo94a}, \cite{AcMo94b}.

\begin{definition}\label{def1}
Let ${\cal A} $ be a $C^*$--algebra.
A {\it localization} in ${\cal A}$, based on the closed intervals of ${R}$
is a two parameter family ${\cal A}_{[s,t]}$ of subalgebras of ${\cal A}$,
such that
\begin{equation}\label{1.9}
[s,t]\subseteq[s',t']\Rightarrow{\cal A}_{[s,t]}\subseteq{\cal
A}_{[s',t']}
\end{equation}
We also require that $\bigcup_{[s,t]} {\cal A}_{[s,t]}$  is dense in
${\cal A}$ and we define ${\cal A}_{t]}$ as a norm clusure of
$\bigcup_{s\leq t} {\cal A}_{[s,t]}$.
\end{definition}

The localization is called expected if, for each
${\cal A}_{t]}$ there is a completely positive norm $1$ projection
({\it Umegaki conditional expectation}), denoted $E_{t]}$, from ${\cal A}$
onto ${\cal A}_{t]}$ satisfying for any $r\le s<t\le u$
$$
E_{s]}E_{t]}= E_{s]} \qquad (\hbox{projectivity})
$$
It is called {\it Markovian} if,
$$E_{t]}{\cal A}_{[t,+\infty)}\subseteq {\cal A}_{t}$$
where ${\cal A}_{t}:={\cal A}_{[t,t]}$.
\medskip

\begin{definition}\label{def2}
Let ${\cal A} $ be a $C^*$--algebra with
a localization ${\cal A}_{[s,t]} $, based on the closed intervals of ${ R}$
and satisfying (\ref{1.9}).
A two parameter family $j_{s,t} \ (s \le t ) \ $ of
maps of ${\cal A}$ into itself satisfying, for every
$r\le s\le t$, the conditions
\begin{equation}\label{1.10}
j_{r,s} \circ j_{s,t} = j_{r,t}
\end{equation}
will be called a {\it right flow (or multiplicative functional)} on ${\cal
A}$.
\end{definition}

If the flow satisfies the identity $j_{s,t}(1)=1$, $\forall s,t$,
then it is called unital (or conservative). We will consider
in this paper only conservative flows.
If the $j_{s,t}$ are completely positive identity preserving and
$$
E_{t]}\circ j_{s,t}= j_{s,t}\circ E_{t]}
$$
\begin{equation}\label{1.11}
j_{s,t} ({\cal A}  _{[r,u]} ) \subseteq {\cal A} _{[r,u]}
\end{equation}
for $r\le s<t\le u$, then the flow is called {\it Markovian},
or a Markov flow.
If on ${\cal A}$ there is a {\it time shift}, i.e. a $1$--parameter
semigroup $u^0_t$ ($t\geq 0$) of left invertible $*$--endomorphisms of
${\cal A} $ satisfying
$$u^0_r{\cal A}_{[s,t]}= {\cal A}_{[s+r,t+r]}$$
$$u^0_rE_{t]}= E_{r+ t]}u^0_r$$
and the flow satisfies the condition
\begin{equation}\label{1.3}
u^0_r \circ j_{s,t} = j_{s+r, t+r} \circ u_r^0
\end{equation}
for all $r,s,t$, then it will be called {\it covariant}. In this case
the one-parameter family
\begin{equation}\label{1.4}
j_t := j_{0,t}
\end{equation}
satisfies the condition
\begin{equation}\label{1.5}
j_{t +s } \circ  u^0_s = j_s \circ u^0_s \circ j_t
\end{equation}
which shows that $j_t $ is a right $(u^0_t)$--{\it cocycle\/} (more
precisely a right $(u^0_t)$--$1$--{\it cocycle }). For this reason a
covariant right Markov flow is also called a {\it right Markov cocycle}.
The quantum Feynman--Kac formula of \cite{Ac78} states that, for
any Markov flow $j_{s,t} $ and for any $s\leq t$, the $2$--parameter family
$$
E_{s]}\circ j_{s,t} \circ u^0_t=:P^{s,t}
$$
is a {\it Markov evolution} on $ {\cal A}$, i.e. the $P^{s,t}$ are
completely positive identity preserving maps of $ {\cal A}$ into itself
satisfying
$$
P^{r,s}P^{s,t}= P^{r,t}\qquad ; \qquad r\leq s\leq t
$$
If the flow $j_{s,t} $ is covariant, then
$$P^{s,t}= P^{0,t-s}=: P^{t-s}\qquad ; \qquad s\leq t$$
and the $1$--parameter family $P^{t}$ is a {\it Markov semigroup}
on $ {\cal A}$.

If all the algebras ${\cal A}_t$ are isomorphic to a single algebra
${\cal B}_S$, and this always happens in the covariant case,
then the evolutions $P^{s,t}$ (respectively the semigroup $P^{t}$)
can all be realized as maps of ${\cal B}_S$ into itself.
In this case the term {\it flow} also for the maps
$j_{s,t}:{\cal B}_S\to {\cal A}$. If
${\cal A}={\cal B}_S\otimes {\cal B}({\cal F})$,
${\cal F}=\Gamma(L^2(R,{\cal K}))$,
which will be the only case considered in our paper starting
from the section (3) on, there is a well known technique
to give a meaning to the flow equation also in this case
\cite{Ac88}, \cite{AcMo94a}.
This technique is discussed in Lemma \ref{lemma9} below (cf.
formula (\ref{2.19})), and includes
the extension of the map
$j_{r,s} : {\cal B}_S \to {\cal B}_S\otimes{\cal B}({\cal F}_{[s,t]})$
to a map from ${\cal B}_S\otimes
{\cal B}({\cal F}_{[s,t]})$ to ${\cal B}({\cal H}_S\otimes{\cal
F}_{[r,t]})$ through the prescription
$$
j_{r,s}(x\otimes X_{s,t})=j_{r,s}(x)\otimes X_{s,t}\equiv
j_{r,s}(x)X_{s,t}
$$
for any $x\in{\cal B}_S$ and $X_{s,t}\in{\cal B}\left({\cal
F}_{[s,t]}\right)$.

\section{Evolutions associated to Markov flows}

In the present paper we consider the flows $j_{s,t}$, where
$j_{s,t}$ are completely positive maps from the $C^*$--algebra
${\cal B}_S={\cal B}({\cal H}_S)$ of all the bounded operators
in the Hilbert space ${\cal H}_S$, called the system space,
with values in the bounded
operators in the Hilbert space ${\cal H}_S\otimes {\cal F}$,
where ${\cal F}=\Gamma\left(L^2(  R,{\cal K})\right)$ is a Bose Fock
space (the reservoir space in physical terminology).

For $f\in L^2(  R,{\cal K})$ the exponential vector $\psi_{f}$
is defined by
$$
\psi_{f}=\sum_{k=0}^{\infty} {1\over\sqrt{k!}} f^{\otimes k}
$$
and enjois the factorization property:
$$
\psi_{f}=\psi_{f_{t]}}\otimes\psi_{f_{[t}}
$$
where $f_{[t,s]}=\chi[t,s]f$ and similarly for $f_{t]}$, $f_{[t}$.

\begin{lemma}\label{lemma3}
Let $j_{s,t}$ be a Markov flow and, for any
pair $f$, $g$ of test functions in $L^2(  R,{\cal K})$ and
$s,t\in  R$, $s\leq t$, define
\begin{equation}\label{2.1}
P^{s,t}_{f,g}(x):=\langle\psi_{f_{[s,t]}},\ j_{s,t}(x)
\psi_{g_{[s,t]}}\rangle;\ x\in{\cal B}_S
\end{equation}
Then each $P^{s,t}_{f,g}$ is a linear map of ${\cal B}_S$ into itself
and
\begin{equation}\label{2.2}
P^{r,t}_{f,g}=P^{r,s}_{f,g}P^{s,t}_{f,g}\ ;\quad
r<s<t
\end{equation}
\end{lemma}

\noindent{\it Proof\/}. Clearly the $P^{s,t}_{f,g}$ map ${\cal B}_S$
into itself. Moreover, in the above notations the
factorization identity for exponential vectors and the
flow equation (\ref{1.10})
imply that for $r<s<t$ and $x\in{\cal B}_S$ one has
$$
P^{r,t}_{f,g}(x)=\langle\psi_{f_{[r,t]}},j_{r,t}(x)\psi_{g_{[r,t]}}
\rangle=\langle\psi_{f_{[r,t]}},j_{r,s}j_{s,t}(x)\psi_{g_{[r,t]}}\rangle=
$$
$$
=\langle\psi_{f_{[r,s]}},j_{r,s}(\langle\psi_{f_{[s,t]}},j_{s,t}
(x)\psi_{g_{[s,t]}}\rangle)\psi_{g_{[r,s]}}\rangle
=P^{r,s}_{f,g}P^{s,t}_{f,g}(x)
$$
Our goal is to reconstruct the flow in terms of the evolutions
$(P^{s,t}_{f,g})$ when $f,g$ vary in a suitably chosen set of test
functions. We shall see that, in the covariant case and
for suitably choosen test functions, the
evolutions $(P^{s,t}_{f,g})$ are in fact semigroups. This will allow to
reduce the theory of flows to the highly developed theory of semigroups.

The semigroups $(P^{t}_{f,g})$ were first introduced by
Fagnola and Sinha \cite{FagnSin} and were extensively used
the papers by Lindsay and Parthasarathy \cite{LiPa} and Lindsay and
Wills \cite{LindWills}, \cite{LindWills1} who gave a new proof, different
from Belavkin's original one \cite{Bel92}, of the characterization, in terms
of structure maps, of completely positive flows, satisfying a stochastic
differential equation. The present paper goes in a different direction,
its main goal being to provide a new infinitesimal characterization of
quantum flows which does not rely on the assumption that the flow satisfies
a stochastic equation. Some recent results of Skeide \cite{Ske2} suggest
that
most of the results of the present paper, at least up to section 5 included,
should continue to hold in the more general framework of tensor product
systems of Hilbert modules.

\medskip

\begin{definition}\label{def4A}
A set ${\cal S}_0\subseteq L^2(  R)$
such that the exponential vectors $\{\psi_f:f\in{\cal S}_0\}$
are total in ${\cal F}=\Gamma(L^2(  R ; {\cal K}))$ is called
{\it totalizing}.
\end{definition}

The following lemma extends a well known property of exponential vectors.
\medskip

\begin{lemma}\label{lemma4}
Let ${\cal S}_0\subseteq L^2(  R)$
be a totalizing set.
Then the set of all linear combinations of the form
\begin{equation}\label{2.3}
\sum_{\alpha\in F}\xi_\alpha\otimes\psi_{f_\alpha}=\psi
\end{equation}
with $F$ a finite set,
$\xi_\alpha\in{\cal H}_S$ and $f_\alpha\in{\cal S}_0$ is a dense
subspace of ${\cal H}_S\otimes{\cal F}$. Moreover the representation
(\ref{2.3}) of a vector $0\not=\psi\in{\cal H}_S\otimes{\cal F}$ is unique
if
the $f_{\alpha}$ are mutually different and
we agree to eliminate from the summation all the $\xi_\alpha$ which are
zero.
We shall denote ${\cal D}({\cal S}_0)$ the subspace of ${\cal
H}_S\otimes{\cal  F}$ of vectors of the form (\ref{2.3}).
\end{lemma}

\noindent{\it Proof\/}. Let ${\cal F}_0$ denote the algebraic linear
span of the vectors $\psi_f$ with $f\in{\cal S}_0$. Then ${\cal
H}_S\otimes{\cal F}_0$ is a dense subspace
of ${\cal H}_S\otimes{\cal F}$ and it is clear that any vector in this
subspace can be written in the form (\ref{2.3}). Suppose now that
\begin{equation}\label{2.4}
\sum_{\alpha\in F}\xi_\alpha\otimes\psi_{f_\alpha}=\sum_{\beta\in G}
\theta_\beta\otimes\psi_{g_\beta}
\end{equation}
are two different representations of a vector $\psi\not=0$.
A vector $\xi\in{\cal
H}_S$, which is orthogonal to all $\xi_\alpha$ will satisfy
$$
\sum_{\beta\in G}\langle\theta_\beta,\xi\rangle\psi_{g_\beta}=0
$$
so it must be also orthogonal to all the $\theta_\beta$. Therefore
we can assume that the $\xi_\alpha$ and the $\theta_\beta$ generate
the same subspace $S_\psi$. If $\xi$ is a non zero vector in this
subspace, then we have
\begin{equation}\label{2.5}
\sum_{\alpha\in F_\xi}\langle\xi_\alpha,\xi\rangle\psi_{f_\alpha}
=\sum_{\beta\in G_\xi}\langle\theta_\beta,\xi\rangle\psi_{g_\beta}
\end{equation}
where $\emptyset\not=F_\xi$ is the set of indices $\alpha$ such that
$\langle\xi_\alpha,\xi\rangle\not=0$ and similarly for $G_\xi$. The
linear independence of the exponential vectors then implies that the
identity (\ref{2.5}) is possible only if the cardinality of $F_\xi$ is equal
to that of $G_\xi$. So up to relabeling the indices we can assume that
$F_\xi=G_\xi$. In this case we must have $\forall\,\alpha\in F_\xi=
G_\xi$:
\begin{equation}\label{2.6}
f_\alpha=g_\alpha\ ,\hbox{ and }\quad
\langle\xi_\alpha,\xi\rangle=\langle\theta_\alpha,\xi\rangle
\end{equation}
Since this must be true for all vectors $\xi$ in the subspace $S_\psi$,
it follows that
\begin{equation}\label{2.7}
\xi_\alpha=\theta_\alpha\ ;\qquad\forall\,\alpha\in F_\xi
\end{equation}
Therefore the identity (\ref{2.4}) is equivalent to
$$\sum_{\alpha\in F\backslash F_\xi}\xi_\alpha\otimes\psi_{f_\alpha}
=\sum_{\beta\in G\backslash F_\xi}\theta_\beta\otimes\psi_{g_\beta}$$
Since $F$ and $G$ are finite sets, iterating this argument we see that
they must have the same cardinality and (up to relabeling) (\ref{2.6})
must hold for all indices $\alpha$.\medskip

\begin{corollary}\label{cor5}
Let ${\cal S}_0$ be as in Lemma \ref{lemma4}. Then
any bounded operator $X\in{\cal B}({\cal H}_S\otimes{\cal F})$ is
uniquely determined by the ${\cal B}({\cal H}_S)$--valued matrix elements
\begin{equation}\label{2.8}
\langle\psi_f,X\psi_g\rangle\quad;\qquad f,g\in{\cal S}_0
\end{equation}
\end{corollary}

\noindent{\it Proof\/}. Let $n\in  N$, $\xi_1,\dots,\xi_n\in{\cal
H}_S$, $f_1,\dots,f_n\in{\cal S}_0$. Then
$$\langle\sum_\alpha\xi_\alpha\otimes\psi_{f_\alpha},\ X\sum_\alpha
\xi_\alpha\otimes\psi_{f_\alpha}\rangle=\sum_{\alpha,\beta}
\langle\xi_\alpha,\langle\psi_{f_\alpha},X\psi_{f_\beta}\rangle
\xi_\beta\rangle$$
So the matrix elements (\ref{2.8}) allow to define all the matrix elements
$\langle\psi,X\psi\rangle$ for $\psi\in{\cal D}({\cal S}_0)$ hence, by
polarization, all the matrix elements $\langle\psi,X\psi'\rangle$ with
$\psi$, $\psi'\in{\cal D}({\cal S}_0)$. Since ${\cal D}({\cal S}_0)$ is
a dense subspace and $X$ is bounded, these matrix elements determine $X$
uniquely.\medskip

\begin{definition}\label{def6}
Let ${\cal S}_0$ be a set and
${\cal A}$, ${\cal B}$ $C^*$--algebra.
A {\it completely positive kernel\/} from ${\cal A}$ to ${\cal B}$,
based on ${\cal S}_0$ is a family
\begin{equation}\label{2.9}
\{P_{f,g}:f,g\in{\cal S}_0\}
\end{equation}
of $C$--linear maps
$P_{f,g}:{\cal A}\to{\cal B}$ such that for any $n\in  N$,
any $f_1,\dots,f_n\in{\cal S}_0$, and any $b_1,\dots,b_n\in{\cal B}$,
the map
\begin{equation}\label{2.10}
x\in{\cal A}\mapsto
\sum^n_{j,k=1}b^*_jP_{f_j,f_k}(x)b_k\in{\cal B}
\end{equation}
is completely positive.
If the maps $P_{f,g}$ are maps from ${\cal B}$ into itself, then
we speak of a completely positive kernel on ${\cal B}$.
\end{definition}
\medskip
\begin{theorem}\label{th7}
Let ${\cal B}_S={\cal B}({\cal H}_S)$
be a $C^*$--algebra, ${\cal S}_0\subseteq L^2 (  R)$
a totalizing set and
$$
P_{f,g}:{\cal B}_S\to{\cal B}_S ;\quad f,g\in{\cal S}_0
$$
a family of linear maps. Then the following are equivalent:
\begin{itemize}
\item{i)} There exists a completely positive map $j:{\cal B}_S\to{\cal B}
({\cal H}_S\otimes{\cal F})$ such that
\begin{equation}\label{2.11}
j(1)=1
\end{equation}
\begin{equation}\label{2.12}
P_{f,g}(x)=\langle\psi_f,j(x)\psi_g\rangle\ ;\quad\forall\,f,g
\in{\cal S}_0\ ,\quad\forall\,x\in{\cal B}_S
\end{equation}
\item{ii)} The family $\{P_{f,g}:f,g\in{\cal S}_0\}$ is a completely
positive kernel on ${\cal B}_S$ based on
${\cal S}_0$ with the property that:
\begin{equation}\label{2.13}
P_{f,g}(1)=e^{\langle f,g\rangle}
\end{equation}
\end{itemize}
\end{theorem}

\noindent{\it Proof\/}. i) $\Rightarrow$ ii). Let $n$, $b_j$, $f_j$ be
as in (\ref{2.10}). Then for any $\xi\in{\cal H}_S$ and $x\in{\cal B}_S$
\begin{equation}\label{2.14}
\langle\xi,\sum_{j,k}b^*_jP_{f_j,f_k}(x)b_k\xi\rangle=
\sum_{j,k}\langle b_j\xi,\langle\psi_{f_j},j(x)\psi_{f_k}\rangle
b_k\xi\rangle=\langle\sum_jb_j\xi\otimes\psi_{f_j},\ j(x)\sum_k b_k
\xi\otimes\psi_{f_k}\rangle
\end{equation}
and, as a function of $x$, the right hand side of (\ref{2.14}) is completely
positive because $j$ has this property. From (\ref{2.12}) with $n=1$,
$b=1_{\cal B}$ using $j(1)=1$, we obtain (\ref{2.13}).

ii) $\Rightarrow$ i). For any $x\in{\cal B}^+_S$ we define a
quadratic form $q_x(\psi,\psi)$ on
the space ${\cal D}({\cal S}_0)$ by:
\begin{equation}\label{2.15}
q_x(\psi,\psi):=\sum_{\alpha,\beta}\langle\xi_\alpha,P_{f_\alpha,
f_\beta}(x)\xi_\beta\rangle
\end{equation}
where $\psi$ is a vector of the form (\ref{2.3}). The complete positivity
property of (\ref{2.10}) implies that, for a positive $x\in {\cal B}_S$
$$
q_x(\psi,\psi)\leq\|x\|q_1(\psi,\psi)=
\|x\|\sum_{\alpha,\beta}\langle\xi_\alpha,\xi_\beta\rangle
e^{\langle f_\alpha,f_\beta\rangle}
=\|x\|\ \left\|\sum\xi_\alpha\otimes\psi_{f_\alpha}\right\|^2=
\|x\|\ \|\psi\|^2$$
This implies that the sesquilinear form,
defined by $q_x(\cdot,\cdot)$ through polarization:
\begin{equation}\label{2.16}
q_x(\psi,\psi')=\sum^3_{n=0}i^nq_x(\psi'+i^n\psi,\psi'+i^n
\psi)
\end{equation}
is continuous and therefore there exists a unique bounded positive
operator $j(x)$ such that
$$
q_x(\psi,\psi)=\langle\psi,j(x)\psi\rangle\ ;\quad\forall\,\psi
\in{\cal D}({\cal S}_0)
$$
and this proves (\ref{2.12}).

The map $x\in{\cal B}^+_S\mapsto j(x)$ is extended to all ${\cal B}_S$ by
complex linearity. We want to prove that this map is completely
positive. To this goal it is sufficient to show that for any vector
$\psi\in{\cal H}_S\otimes{\cal F}$, for any $n\in  N$ and for any
$a_1,\dots,a_n\in{\cal B}({\cal H}_S\otimes{\cal F})$
and $x_1,\dots,x_n\in{\cal B}_S$, one has
\begin{equation}\label{2.17}
\sum_{j,k=1}^n\langle\psi,a^*_jj(x^*_jx_k)a_k\psi\rangle=
\sum_{j,k=1}^n\langle a_j\psi,j(x^*_jx_k)a_k\psi\rangle\geq0
\end{equation}
Since the vectors of the form (\ref{2.3}) are dense in
${\cal H}_S\otimes{\cal F}$,
for each $j=1,\dots,n$ there is a sequence
$\left(\sum_{\alpha \in
F_{j,m}}\xi^{(m)}_{j,\alpha}\otimes\psi_{f^{(m)}_{j,\alpha}} \right)_m$,
where $F_{j,m}$ is a finite set, $\xi^{(m)}_{j,\alpha} \in{\cal H}_S$ and
$f^{(m)}_{j,\alpha}\in{\cal S}_0$, such that
$$a_j\psi=\lim_{m\to+\infty}\sum_{\alpha\in F_{j,m}}\xi^{(m)}_{j,
\alpha}\otimes\psi_{f^{(m)}_{j,\alpha}}$$

Moreover, since $j$ runs over the finite set
$1,\dots,n<+\infty$, then, possibly by defining some
$\xi^{(m)}_{j,\alpha}$ to be equal to zero, we can suppose that the
index set $F_{j,m}$ does not depend on $j$, i.e.
$$F_{j,m}=F_m\quad\hbox{(finite set) };\quad\forall\,j=1,\dots,n;\
\forall\,m$$
With this convention, the left hand side of (\ref{2.17}) is equal to
$$
\lim_{m\to\infty}\sum_{j,k=1}^n\sum_{\alpha\in F_m}\sum_{\beta\in F_m}
\langle\xi^{(m)}_{j,\alpha}\otimes\psi_{f^{(m)}_{j,\alpha}},
j_{s,t}(x^*_jx_k)\xi^{(m)}_{k,\beta}\otimes\psi_{f^{(m)}_{k,\beta}}
\rangle=
$$
$$
=\lim_{m\to\infty}\sum_{(j,\alpha)\in\{1,\dots,n\}\times F_m}\sum_{(k,\beta)
\in\{1,\dots,n\}\times F_m}\langle\xi^{(m)}_{j,\alpha},
P_{f^{(m)}_{j,\alpha},f^{(m)}_{k,\beta}}(x^*_jx_k)\xi^{(m)}_{k,\beta}\rangle
$$
which is $\geq0$ because of the complete positivity property (\ref{2.10}).

Finally, having proved (\ref{2.17}), (\ref{2.12}) follows from
(\ref{2.15}) and Corollary \ref{cor5}.
\medskip

\begin{corollary}\label{cor8}
Let $s,t\in  R$ with $s<t$ and let
${\cal S}_{0,[s,t]}\subseteq L^2(  R)$ be a set of functions
with support in $[s,t]$ and totalizing for
${\cal F}_{[s,t]}$. Let ${\cal B}_S$ and
$\{P_{f,g}:f,g\in{\cal S}_{0,[s,t]}\}$ be as in
Theorem \ref{th7}. Then the following are equivalent:
\begin{itemize}
\item{(i)} There exists a completely positive map
$j_{s,t}:{\cal B}_S\to
{\cal B}({\cal H}_S\otimes{\cal F}_{[s,t]})$ such that
(\ref{2.11}) and (\ref{2.12}) hold.
\item{(ii)} Condition (ii) of Theorem \ref{th7} is satisfied for all
$f,g\in{\cal S}_{0,[s,t]}$.
\end{itemize}
\end{corollary}

\noindent{\it Proof\/}. This is obtained from the proof of Theorem \ref{th7}
replacing everywhere ${\cal F}$ by ${\cal F}_{[s,t]}$.
\medskip

A known theorem by Schur states that, if $a=(a_{ij})$ and $b=(b_{ij})$
are positive definite matrices then their pointwise product $c_{ij}=
a_{ij}b_{ij}$ is positive definite. The following is a generalization
of this result to completely positive kernels.\medskip

\begin{lemma}\label{lemma9}
In the notations and assumptions of Theorem \ref{th7}
let  $(Q_{f,g})$ and $(P_{f,g})$ be completely positive kernels on
${\cal B}_S$ based on $\chi_{[r,s]}{\cal S}_0$
and $\chi_{[s,t]}{\cal S}_0$ respectively, we use the convention
$$
Q_{f,g}=Q_{\chi_{[r,s]}f,\chi_{[r,s]}g},\qquad
P_{f,g}=P_{\chi_{[s,t]}f,\chi_{[s,t]}g}.
$$
Then their product
\begin{equation}\label{2.18}
Q_{f,g}P_{f,g}\quad;\qquad f,g\in{\cal S}_0
\end{equation}
meant in the sense of composition of maps from ${\cal B}_S$ to ${\cal
B}_S$ is also a completely positive kernel.
\end{lemma}

\noindent{\it Proof\/}. Fix $r,s,t\in  R$, with $r<s<t$ and let
$j_{r,s}$ denote the completely positive map from ${\cal B}_S$ to ${\cal
B}({\cal H}_S\otimes{\cal F}_{[r,s]})$ associated to the completely
positive kernel $(Q_{f,g})$ with $f$, $g\in\chi_{[r,s]}{\cal S}_0$. Let
$j_{s,t}$ denote the completely positive map from ${\cal B}_S$ to ${\cal
B}({\cal H}_S\otimes{\cal F}_{[s,t]})$ obtained in the same way from
$(P_{f,g})$, with $f$, $g\in\chi_{[s,t]}{\cal S}_0$.
Now we extend the map $j_{r,s}$ to a map from ${\cal B}_S\otimes
{\cal B}({\cal F}_{[s,t]})$ to ${\cal B}({\cal H}_S\otimes{\cal
F}_{[r,t]})$ by the prescription
\begin{equation}\label{2.19}
j_{r,s}(x\otimes X_{s,t})=j_{r,s}(x)\otimes X_{s,t}\equiv j_{r,s}
(x)X_{s,t}
\end{equation}
for any $x\in{\cal B}_S$ and $X_{s,t}\in{\cal B}\left({\cal
F}_{[s,t]}\right)$. This extension is completely positive being identified
to $j_{r,s}\otimes id_{s,t}$. Notice that (\ref{2.19}) implies that
\begin{equation}\label{2.20}
\langle\psi_f,j_{r,s}(A_{s,t})\psi_g\rangle:=
\langle\psi_{f_{[r,s]}},j_{r,s}(\langle\psi_{f_{[s,t]}},
A_{s,t}\psi_{g_{[s,t]}}\rangle)\psi_{g_{[r,s]}}\rangle
\end{equation}
for any operator $A_{s,t}\in{\cal B}_S\otimes{\cal B}({\cal F}_{[s,t]})$
and for any $f$, $g\in\chi_{[r,t]}{\cal S}_0$.

Using (\ref{2.19}) and (\ref{2.20}) we obtain,
for any $x\in{\cal B}_S$ and $f$,
$g\in\chi_{[r,t]}{\cal S}_0$:
\begin{equation}\label{2.21}
\langle\psi_f,j_{r,s}j_{s,t}(x)\psi_g\rangle=
\langle\psi_{f_{[r,s]}},j_{r,s}(\langle\psi_{f_{[s,t]}},j_{s,t}
(x)\psi_{g_{[s,t]}}\rangle)\psi_{g_{r,s]}}\rangle=
Q_{f,g}P_{f,g}(x)
\end{equation}
But $j_{r,s}\circ j_{s,t}:{\cal B}_S\to{\cal B}({\cal H}_S
\otimes{\cal F}_{[r,t]})$ is a completely positive map and therefore the
left hand side of (\ref{2.21})
is a completely positive kernel on ${\cal B}_S$ based
on $\chi_{[r,t]}{\cal S}_0$. This proves the lemma.\medskip

\begin{theorem}\label{th12}
Let $j_{s,t}$ be a Markov flow on
$C^*$--algebra ${\cal A}={\cal B}\left({\cal H}_S\otimes{\cal F}\right)$
with the localization
${\cal A}_{[s,t]}={\cal B}\left({\cal H}_S\otimes{\cal F}_{[s,t]}\right)$
and let ${\cal S}_0 \subseteq L^2(  R)$
be a totalizing set. Then the family
\begin{equation}\label{2.23}
\{P^{s,t}_{f,g}:s,t\in  R\ ;\quad s<t\ ;\quad f,g\in{\cal S}_0\}
\end{equation}
defined by (\ref{2.1}) has the following properties:
\begin{itemize}
\item{i)} for any $f$, $g\in{\cal S}_0$ and for any $r,s,t\in  R$
with $r<s<t$, $P^{s,t}_{f,g}$ is a linear map of
${\cal B}_S={\cal B}({\cal H}_S)$ into itself and:
\begin{equation}\label{2.24}
P^{r,t}_{f,g}=P^{r,s}_{f,g}P^{s,t}_{f,g}
\end{equation}
\item{ii)} for any $f$, $g\in{\cal S}_0$ and $s$, $t\in  R$, $s<t$
\begin{equation}\label{2.25}
P^{s,t}_{f,g}(1)=e^{\langle\chi_{[s,t]}f,\chi_{[s,t]}g\rangle}
\end{equation}
\item{iii)} for each $s,t\in  R$ with $s<t$, the family
$\{P^{s,t}_{f,g}:f,g\in{\cal S}_0\}$ is a completely positive kernel
on ${\cal B}_S$ based on ${\cal S}_0$.
\end{itemize}

Conversely, given a family of the form (\ref{2.23})
satisfying (i), (ii), (iii),
there exists a conservative Markov flow $j_{s,t}$ on ${\cal A}$
such that each $P^{s,t}_{f,g}$ is given by formula (\ref{2.1}).
\end{theorem}

\noindent{\it Proof\/}. Necessity. If $j_{s,t}$ is a Markov flow,
then properties (i) and (ii) follow from Lemma \ref{lemma3}
and property (iii) from Theorem \ref{th7}.

Sufficiency. Let $(P^{s,t}_{f,g})$ be a family satisfying (i), (ii),
(iii). Then, for any $f$, $g\in{\cal S}_0$ and $s,t\in  R$, $s<t$, we
know from Corollary \ref{cor8} that there exists a linear, completely
positive, identity preserving map
$$
j_{s,t}:{\cal B}_S\to{\cal B}({\cal
H}_S\otimes{\cal F}_{[s,t]})$$ characterized by
$$\langle\psi_f,j_{s,t}(x)\psi_g\rangle=P^{s,t}_{f,g}(x)\ ;\quad
\forall\,x\in{\cal B}_S
$$
for any $f$, $g\in{\cal S}_0$ with supp $f$, supp $g\subseteq[s,t]$.
Each $j_{s,t}(x)$ $(x\in{\cal B}_S)$ is then uniquely extended to an
operator in ${\cal B}({\cal H}_S\otimes{\cal F})$, still denoted with
the same symbol, by the prescription
\begin{equation}\label{2.26}
\langle\psi_f,j_{s,t}(x)\psi_g\rangle:=
\langle\psi_{\chi_{[s,t]^c f}},
\psi_{\chi_{[s,t]^c g}}\rangle P^{s,t}_{f,g}(x)
\end{equation}
where $\chi_{[s,t]^c}=1-\chi_{[s,t]}$. We now extend the map
$$
j_{s,t}:{\cal B}_S\to {\cal B}({\cal H}_S\otimes{\cal F}_{[s,t]})
$$
to a map
$$
j_{s,t}:{\cal B}_S\otimes{\cal B}({\cal F}_{[t})\to{\cal B}({\cal H}_S
\otimes{\cal F}_{[s})
$$
by the prescription
$$
\langle\psi_f,j_{s,t}(X_{[t})\psi_f\rangle:=\langle\psi_{f_{t]}},
j_{s,t}(\langle\psi_{f_{[t}},X_{[t}\psi_{g_{[t}}\rangle)
\psi_{g_{t]}}\rangle$$
for any $f$, $g\in{\cal D}_0$ and any operator $X_{[t}$ in ${\cal
B}({\cal H}_S\otimes{\cal F}_{[t})$.
With this prescription it makes sense to speak of $j_{r,s}j_{s,t}(x)$,
for $x\in{\cal B}_S$ and $r<s<t$. Moreover one has, for any
$f,g\in{\cal D}_0$:
$$
\langle\psi_f,j_{r,s}j_{s,t}(x),\psi_g\rangle=
\langle\psi_{f_{s]}},j_{r,s}(\langle\psi_{f[s,t]},j_{s,t}(x)
\psi_{g_{[s,t]}}\rangle)\psi_{g_{s]}}\rangle=
$$
$$
=\langle\psi_{f_{[r,s]}},j_{r,s}(P^{s,t}_{f,g}(x))\psi_{g_{[r,s]}}
\rangle\langle\psi_{\chi_{[r}f},\psi_{\chi_{[r}g}\rangle=
$$
$$
=P^{r,s}_{f,g}P^{s,t}_{f,g}(x)\langle\psi_{\chi_{[r,t]}^cf},
\psi_{\chi_{[r,t]^c}g}\rangle
=P^{r,t}_{f,g}(x)\langle\psi_{\chi_{[r,t]^c}f},\psi_{\chi_{[r,t]^c}g}
\rangle=\langle\psi_f,j_{r,t}(x)\psi_g\rangle
$$
Therefore $j_{s,t}$ is a right multiplicative functional.

%We define a conditional expectation $E_{t]}$ as a map in
%${\cal B}({\cal H}_S\otimes {\cal F})$ defined as follows
%$$
%E_{t]}(X)=\Pi_{t]}X\Pi_{t]}
%$$
%where $\Pi_{t]}$ is an orthogonal projection on the subspace,
%generated by $1_{{\cal H}_S}\otimes\psi_{f_{t]}}$.
Because of (\ref{2.26}) $j_{s,t}$ is localized in $[s,t]$,
therefore $j_{s,t}$ is a Markov flow.

\section{Semigroups associated to Markovian cocycles}

\begin{lemma}\label{lemma13}
Suppose that $f$, $g\in L^2(  R,{\cal K})$
assume a constant value, in the interval $[s,t]$, equal respectively to
$f_0$ and $g_0$ (vectors in ${\cal K}$).
Then if $j_{s,t}$ is a covariant Markovian cocycle, for
any $x\in{\cal B}_S$, one has
\begin{equation}\label{3.1}
\langle\psi_{f_{[s,t]}},j_{s,t}(x)\psi_{g_{[s,t]}}\rangle=
P^{s,t}_{f,g}(x)=
\langle\psi_{\chi_{[0,t-s]}f_0},j_{0,t-s}(x)\psi_{\chi_{[0,t-s]}
g_0}\rangle
\end{equation}
\end{lemma}

\noindent{\it Proof\/}. The covariance condition (\ref{1.3}) implies that
$$\langle\psi_{f_{[s,t]}},j_{s,t}(x)\psi_{g_{[s,t]}}\rangle=
\langle\psi_{f_{[s,t]}},u^0_sj_{0,t-s}(x)\psi_{g_{[s,t]}}\rangle$$
using the explicit form of $u^0_t$
$$u^0_s(x) = \Gamma(S_s)(x)\Gamma(S_s)^*$$
where $S_s$ is the shift in $L^2(  R,{\cal K})$, defined by
$$S_{s}f(\tau)=f(\tau -s)$$
this becomes
$$\langle\psi_{f_{[s,t]}},\Gamma(S_s)j_{0,t-s}(x)\Gamma(S_s)^*
\psi_{g_{[s,t]}}\rangle
=\langle\Gamma(S^*_s)\psi_{f_{[s,t]}},j_{0,t-s}(x)\Gamma(S^*_s)
\psi_{g_{[s,t]}}\rangle=$$
\begin{equation}\label{3.2}
=\langle\psi_{S_{-s}f_{[s,t]}},j_{0,t-s}(x)\psi_{S_{-s}g_{[s,t]}}
\rangle
\end{equation}
Under our assumptions on $f$ and in the notation (x.), one has, for
$\tau\in[0,t-s]$:
$$S_{-s}f_{[s,t]}(\tau)=S_{-s}\chi_{[s,t]}f(\tau)=\chi_{[s,t]}
(\tau+s)f(\tau+s)=\chi_{[0,t-s]}(\tau_0)f_0$$
Therefore the right hand side of (\ref{3.2}) is equal to
$$\langle\psi_{f_0\chi_{[0,t-s]}},j_{0,t-s}(x)\psi_{g_0\chi_{[0,t-s]}}
\rangle$$
and this proves (\ref{3.1}).

\begin{lemma}\label{lemma14}
Let $f$, $g$ be as in Lemma \ref{lemma13} and define, for
$\tau\in[0,b-a]$ and $x\in{\cal B}_S$:
\begin{equation}\label{3.3}
P^\tau_{f,g}(x):=\langle\psi_{\chi_{[0,\tau]}f_0},j_{0,\tau}
(x)\psi_{\chi_{[0,\tau]}g_0}\rangle
\end{equation}
Then $P^\tau_{f,g}:{\cal B}_S\to{\cal B}_S$ is the restriction of a
semigroup to the interval $[0,b-a]$, i.e. if $\rho$, $\sigma\in[0,
b-a]$ are such that $\rho+\sigma\in[0,b-a]$, then
$$P^\rho_{f,g}P^\sigma_{f,g}=P^{\rho+\sigma}_{f,g}
$$
\end{lemma}

\begin{remark}\label{rem15}
If a semigroup $(P^t)$ is defined on an
interval $[0,T]$ one can always extend it to $[0,2T]$ by putting
$$P^{T+s}:=P^TP^s$$
therefore, proceeding by induction, one can extend it to the whole of
$  R_+$.
Clearly if $P^t$ is strongly continuous in $[0,T]$ its extension will
have the same property.
\end{remark}

\noindent{\it Proof\/}. It is convenient to write
$$\rho=s-r\quad;\qquad\sigma=t-s$$
with $r$, $s$, $t\in[a,b]$. Then we have
\begin{equation}\label{3.4}
P^{s-r}_{f,g}P^{t-s}_{f,g}(x)=
\langle\psi_{\chi_{[0,s-r]}f_0},j_{0,s-r}(\langle\psi_{\chi_{[0,t-
s]}f_0},j_{0,t-s}(x)\psi_{\chi_{[0,t-s]}g_0}\rangle)\psi_{\chi_{[0,
s-r]}g_0}\rangle
\end{equation}
Using the identity (\ref{3.1}) the right hand side of (\ref{3.4}) becomes
$$
\langle\psi_{\chi_{[a+r,a+s]}f_0},j_{a+r,a+s}(\langle\psi_{\chi_{[a+
s,a+t]}f_0},\
j_{a+s,a+t}(x)\psi_{\chi_{[a+s,a+t]}g_0}\rangle)\psi_{\chi_{[a+r,
a+s]}g_0}\rangle
$$
and, because of the factorization properties of the exponential vectors,
this is equal to
$$\langle\psi_{\chi_{[a+r,a+t]}f_0},j_{a+r,a+s}j_{a+s,a+t}(x)
\psi_{\chi_{[a+r,a+t]}g_0}\rangle=
\langle\psi_{\chi_{[a+r,a+t]}f_0},j_{a+r,a+t}(x)\psi_{\chi_{[a+r,
a+t]}g_0}\rangle$$
Using again the identity (\ref{3.1}), this is equal to
$$\langle\psi_{\chi_{[0,t-r]}f_0},j_{0,t-r}(x)\psi_{\chi_{[0,t-r]}
g_0}\rangle=P^{t-r}_{f,g}(x)$$
and this ends the proof.\medskip

\begin{lemma}\label{lemma17}
In the notations and assumptions of Lemma \ref{lemma13},
if the cocycle $(j_{s,t})$ is strongly continuous, then the semigroups
$P^\tau_{f,g}$, defined by (\ref{3.3}) are strongly continuous and one
has, denoting $|f|$ the norm in ${\cal K}$):
\begin{equation}\label{3.5}
\|P^t_{f,g}(x)\|\leq
e^{{t\over2}\,(|f_0|^2+|g_0|^2)}\|x\|
\end{equation}
\end{lemma}

\noindent{\it Proof\/}.\qquad From (\ref{3.3}) we deduce
$$
\|P^t_{f,g}(x)\|\leq\|\psi_{\chi_{[0,t]}f_0}\|\cdot\|x\|\cdot
\|\psi_{\chi_{[0,t]}g_0}\|=e^{-{t\over2}\,(|f_0|^2+|g_0|^2)}\|x\|
$$
The strong continuity is clear.\medskip

\begin{theorem}\label{th18}
Let ${\cal F}=\Gamma(L^2(  R))$ and let
$(j_{s,t})$ be a Markov flow from ${\cal B}_S$ to ${\cal B}({\cal H}_S
\otimes{\cal F})$. Define, for any $t\geq0$ and any pair of vectors
$f_0, g_0\in {\cal K}$ the map
\begin{equation}\label{3.6}
P^t_{f_0,g_0}(x):=\langle\psi_{f_0\chi_{[0,t]}},j_{0,t}(x)
\psi_{g_0\chi_{[0,t]}}\rangle\ ;\quad x\in{\cal B}_S
\end{equation}
Then:
\begin{itemize}
\item{(i)} for any $t\geq0$, $f_0$, $g_0\in {\cal K}$, $P^t_{f_0,g_0}$ is a
$C_0$--semigroup from ${\cal B}_S$ into itself
\item{(ii)} for $t$, $f_0$, $g_0$ as in (i)
\begin{equation}\label{3.7}
P^t_{f_0,g_0}(1)=e^{t \langle f_0,g_0\rangle_{\cal K} }
\end{equation}
\item{(iii)} for each $t\geq0$, the family
\begin{equation}\label{3.8}
\{P^t_{f_0,g_0}:f_0,g_0\in {\cal K}  \}
\end{equation}
is a completely positive kernel from ${\cal B}_S$ to ${\cal B}_S$ based
on ${\cal K} $.
\end{itemize}

Conversely, given a family of the form (\ref{3.8}) satisfying (i), (ii) and
(iii), there exists a conservative Markov flow from ${\cal B}_S$
to ${\cal B}({\cal H}_S\otimes{\cal F})$ such that each
$P^t_{f_0,g_0}$ is given by (\ref{3.6}).
\end{theorem}

\noindent{\it Proof\/}. It is a known and elementary fact that the
family ${\cal S}_0$ of functions of the form $f_0\chi_{[0,t]}$, with
$f_0\in{\cal K}$ and $\chi_{[0,t]}$ a characteristic function of a
bounded interval in $  R$, is totalizing for ${\cal F}$. With
this choice of ${\cal S}_0$ we can apply Theorem \ref{th12} and conclude
that
the assignment of a Markov flow is equivalent to the assignment of a
family $(P^{s,t}_{f,g})$ with $s$, $t\in  R$, $s\leq t$, $f$,
$g\in{\cal S}_0$.
Now fix, $s$, $t$, $f$, $g$ as above. Then there exists a partition
$$s=t_0<t_1<\dots<t_n<t=t_{n+1}$$
of the interval $[s,t]$ such that $f$ (resp. $g$) has the constant value
$f_j\in  C$ (resp. $g_j\in  C$) in the interval $[t_j,t_{j+1})$
$(j=0,\dots,n)$. Correspondingly we have
\begin{equation}\label{3.9}
P^{s,t}_{f,g}=P^{s,t_1}_{f,g}P^{t_1,t_2}_{f,g}\cdot\dots\cdot
P^{t_n,t}_{f,g}=P^{t_1-s}_{f_0,g_0}P^{t_2-t_1}_{f_1,g_1}\dots
P^{t-t_n}_{f_{n+1},g_{n+1}}
\end{equation}
Therefore the assignment of the family $(P^{s,t}_{f,g})$, or
equivalently of the flow, is equivalent to the assignment of the family
of semigroups $(P^t_{f_0,g_0})$ $(t\geq0,f_0,g_0\in  C)$. Moreover,
by the non commutative Schur Lemma \ref{lemma9},
if for each $t\in  R_t$ the
maps $\{P^t_{f_0,g_0}:f_0,g_0\in  C\}$ are a completely positive
kernel, then for each $s\leq t$ the maps $\{P^{s,t}_{f,g}:f,g
\in{\cal S}_0\}$ have the same property. Since the converse is clear
because the semigroups $P^t_{f_0,g_0}$ are a subset of the maps
$P^{s,t}_{f,g}$, the theorem is proved.

\section{The extended semigroup of a flow}

In this section restrict our considerations to the case ${\cal K}=   C$,
so that $L^2(  R, {\cal K})= L^2(  R) $. We shall use the following

\begin{theorem}\label{th19}
Let ${\cal S}_0\subseteq L^2(  R)$ denote
the set of finite sums of characteristic functions over bounded disjoint
intervals, i.e.
\begin{equation}\label{4.1}
{\cal S}_0=\left\{\sum^n_{j=1}\chi_{[a_j,b_j]};\ n\in  N,\
a_i,b_j\in  R,\ a_j<b_j,\ (a_j,b_j)\cap(a_k,b_k)=\emptyset,\hbox{ if }
j\not=k\right\}
\end{equation}
Then the set of exponential vectors with test functions in ${\cal S}_0$
are total in $L^2(  R)$.
\end{theorem}

\noindent{\it Proof\/}. An elementary proof of this theorem is in
\cite{Ske99}.
More elaborated proofs are in \cite{PaSu98}, \cite{Bha98}.
\medskip
In this section the family ${\cal S}_0$ will be fixed and given by
(\ref{4.1}).
\medskip
\begin{theorem}\label{th20}
Let ${\cal F}=\Gamma(L^2(  R))$ and let
$(j_{s,t})$ be a Markov flow from ${\cal B}_S$ to ${\cal B}({\cal H}_S
\otimes{\cal F})$. Let $\psi_0$ denote the vacuum vector in ${\cal F}^0$
(we use the same notation for the vacuum vector in any ${\cal
F}_{[s,t]}$).
Define, for $x\in{\cal B}_S$ and $t\geq0$
\begin{equation}\label{4.2}
P^t_{00}(x):=\langle\psi_0,j_{0,t}(x)\psi_0\rangle
\end{equation}
\begin{equation}\label{4.3}
P^t_{01}(x):=\langle\psi_0,j_{0,t}(x)\psi_{\chi_{[0,t]}}\rangle
\end{equation}
\begin{equation}\label{4.4}
P^t_{10}(x):=\langle\psi_{\chi_{[0,t]}},j_{0,t}(x)\psi_0\rangle
\end{equation}
\begin{equation}\label{4.5}
P^t_{11}(x):=\langle\psi_{\chi_{[0,t]}},j_{0,t}(x)\psi_{\chi_{[0,t]}}\rangle
\end{equation}
Then:
\begin{itemize}
\item{i)} for each $\varepsilon$, $\varepsilon'\in\{0,1\}$,
$(P^t_{\varepsilon,\varepsilon'})$ is a $C_0$--semigroup on ${\cal B}_S$
\item{ii)} the semigroups $(P^t_{\varepsilon,\varepsilon'})$ satisfy:
\begin{equation}\label{4.6}
P^t_{00}(1)=P^t_{10}(1)=P^t_{01}(1)=1
\end{equation}
\begin{equation}\label{4.7}
P^t_{11}(1)=e^t
\end{equation}
\item{iii)} the map:
\begin{equation}\label{4.8}
x\in{\cal B}_S\mapsto\pmatrix{
P^t_{00}(x)&P^t_{01}(x)\cr
P^t_{10}(x)&P^t_{11}(x)\cr}\in M(2,C)\otimes{\cal B}_S
\end{equation}
where $M(2,C)$ denotes the algebra of $2\times2$ complex matrices, is
completely positive.
\end{itemize}

Conversely, given four $C_0$--semigroups $(P^t_{\varepsilon,
\varepsilon'})$ $(\varepsilon,\varepsilon'\in\{0,1\})$ satisfying
condition (\ref{4.6}), (\ref{4.7}), (\ref{4.8}) then there exists a unique
conservative Markov flow on ${\cal B}_S\otimes{\cal B}({\cal F})$
satisfying conditions (\ref{4.2}), (\ref{4.3}), (\ref{4.4}), (\ref{4.5}).
\end{theorem}

\noindent{\it Proof\/}. Let $f$, $g\in{\cal S}_0$ and $s$, $t\in  R$,
$s<t$. Then there exists a partition
\begin{equation}\label{4.9}
s=t_0<t_1<\dots<t_n<t_{n+1}=t
\end{equation}
such that both $f$ and $g$ are constant in each interval of this
partition.
In our assumptions this constant value can only be 0 or 1 so, in the
notation (\ref{3.6}), we have only four possibilies:
\begin{equation}\label{4.10}
P^t_{00},P^t_{01},P^t_{10},P^t_{11}
\end{equation}
In these notations
\begin{equation}\label{4.11}
P^{s,t}_{f,g}=P^{t_1,s}_{\varepsilon_0,\delta_0}P^{t_2-t_1}_{\varepsilon_1,
\delta_1}\dots P^{t-t_n}_{\varepsilon_{n+1},\delta_{n+1}}
\end{equation}
where $\varepsilon_j$, $\delta_j\in\{0,1\}$. Thus $(P^{s,t}_{f,g})$ is
uniquely determined by the four semigroups (\ref{4.2}),..., (\ref{4.5}).

According to Definition \ref{def6} the complete positivity of the kernel
$(P^t_{f_0,g_0})$ means that, for any $n\in  N$,
$b_1,\dots,b_n\in{\cal B}_S$, $f_1,\dots,f_n\in  C$ the map
\begin{equation}\label{4.12}
x\in{\cal B}_S\mapsto\sum^n_{j,k=1}b^*_jP^t_{f_j,f_k}(x)b_k
\end{equation}
is completely positive. If the $f_j$--can only take values 0 and 1 we
can assume, up to a relabeling of the indices, that
\begin{equation}\label{4.13}
f_j=0\hbox{ for }j=1,\dots,n_1\ ;\quad f_j=1\hbox{ for }
j=n_1+1,\dots,n
\end{equation}
With this notation the right hand side of (\ref{4.12}) becomes
$$\sum^{n_1}_{j,k=1}b^*_jP^t_{00}(x)b_k+\sum^{n_1}_{j=1}
\sum^n_{k=n_1+1}b^*_jP^t_{01}(x)b_k
+\sum^n_{j=n_1+1}\sum^{n_1}_{k=1}b^*_jP^t_{10}(x)b_k+
\sum^n_{j,k=n_1+1}b^*_jP^t_{11}(x)b_k=$$
$$=c^*_0P^t_{00}(x)c_0+c^*_0P^t_{01}(x)c_1+
c^*_1P^t_{10}(x)c_0+c^*_1P^t_{11}(x)c_1$$
where we have put
$$c_0:=\sum^{n_1}_{j=1}b_j\quad;\qquad c_1:=\sum^n_{j=n_1+1}b_j$$
Since the $b_j$ are arbitrary in ${\cal B}_S$ so are $c_0$, $c_1$. So,
under our assumptions, the complete positivity of the map (\ref{4.12}) is
equivalent to the complete positivity of the map (\ref{4.8}).

Finally from condition (ii) of Theorem \ref{th18} one immediately deduces
(\ref{4.6}), (\ref{4.7}).

Conversely, let be given the four semigroups
(\ref{4.2}), \dots, (\ref{4.5}). Then, for $f$,
$g\in{\cal S}_0$ with associated partition (\ref{4.9}), we can define
$P^{s,t}_{f,g}$ by (11). The evolution property
(\ref{3.2}) follow immediately from
the semigroup property and (\ref{4.11}).

Because of complete positivity of
(\ref{4.8}), $\{P^{s,t}_{f,g}:f,g\in{\cal S}_0\}$,
defined by (\ref{4.11}), is a
completely positive kernel on ${\cal B}_S$ based on ${\cal S}_0$.
This finishes the proof of the theorem.
\medskip

Formula (\ref{4.8}) naturally suggests to study the following semigroup:
\medskip

\begin{definition}\label{def_extsgr}
The $1$--parameter semigroup
\begin{equation}\label{extsgr}
\pmatrix{
x_{00}&x_{11}\cr
x_{10}&x_{11}\cr}\in M(2,C)\otimes{\cal B}_S\mapsto\pmatrix{
P^t_{00}(x_{00})&P^t_{01}(x_{01})\cr
P^t_{10}(x_{10})&P^t_{11}(x_{11})\cr}\in M(2,C)\otimes{\cal
B}_S
\end{equation}
will be called {\it the extended semigroup\/} of the flow $(j_{s,t})$
and denoted $\tilde P^t$.
\end{definition}

It is clear that the map (\ref{4.8}) is obtained by  restriction of the
extended semigroup $\tilde P^t$ to the subspace of $M_2\otimes{\cal B}_S$
formed by the matrices of the form
$$
\pmatrix{
x&x\cr
x&x\cr}\qquad ; \qquad x\in{\cal B}_S
$$

\begin{lemma}\label{semi_n}
Let ${\cal H}_S$ and ${\cal F}$ be Hilbert
spaces and for $n\in  N$, let be given $n$ vectors
$\psi_1,\dots,\psi_n\in{\cal F}$. Then the map
$$
X=(x_{jk})\in M(n,{\cal B}({\cal H}_S\otimes{\cal F}))\mapsto(
\langle\psi_j,x_{jk}\psi_k\rangle)\in M(n,{\cal B}({\cal H}_S))
$$
is completely positive.
\end{lemma}

\noindent{\it Proof\/}. It is sufficient to prove the positivity of the
above map because, since ${\cal H}_S$ is arbitrary, we can always
replace it by ${\cal H}_S\otimes  C^k$ $(k\in  N)$.
Let $x=(x_{jk})\in M(n,{\cal B}({\cal H}_S\otimes{\cal F}))$ be
positive. We want to show that, for any $c_1,\dots,c_n\in{\cal B}({\cal
H}_S)$ the operator
\begin{equation}\label{map_n}
\sum_{jk}c^*_j\langle\psi_j,x_{jk}\psi_k\rangle c_k\in{\cal B}
({\cal H}_S)
\end{equation}
is positive. Clearly (\ref{map_n}) is a self--adjoint element of
${\cal B}({\cal H}_S)$ and, if
$\xi\in{\cal H}_S$ is any vector, then
$$\langle\xi,\sum_{jk}c^*_j\langle\psi_j,x_{jk}\psi_k\rangle c_k
\xi\rangle=\sum_{jk}\langle c_j\xi,\langle\psi_j,x_{jk}\psi_k
\rangle c_k\xi\rangle=$$
$$=\sum_{jk}\langle c_j\xi\otimes\psi_j,x_{jk}(c_k\xi\otimes\psi_k)
\rangle=\langle\varphi,x\varphi\rangle\geq0
$$
where $\varphi:=\otimes^n_{j=1}c_j\xi\otimes\psi_j$.

\begin{theorem}\label{ext_semi}
The extended semigroup
$\tilde P^t:M(2,{\cal B}_S)\to M(2,{\cal B}_S)$
is a completely positive semigroup satisfying the condition
\begin{equation}\label{norm}
\tilde P^t\pmatrix{
1&1\cr
1&1\cr}=\pmatrix{
1&1\cr
1&e^t\cr}
\end{equation}
Conversely any completely positive semigroup $\tilde P^t$ on
$M(2, {\cal B}_S)$ satisfying condition (\ref{norm})
is the extended semigroup of a unique Markov flow on ${\cal B}_S$
which is completely determined by the coefficients of $\tilde P^t$
via Theorem \ref{th20}.
\end{theorem}

\noindent{\it Proof\/}. {\it Necessity\/}. The complete positivity of
$(j_{s,t})$ implies that the flow
$$
j^{(2)}_{s,t}:(x_{ij})\in M(2,{\cal B}_S)\to(j_{s,t}(x_{ij}))
\in M(2,{\cal B} \left({\cal H}_S\otimes{\cal F})\right)
$$
is completely positive. Since $\tilde P^t$ is obtained by composing this
flow with the map
$$\pmatrix{
j_{0,t}(x_{00})&j_{0,t}(x_{01})\cr
j_{0,t}(x_{10})&j_{0,t}(x_{11})\cr} \mapsto
\pmatrix{
\langle \psi_0,j_{0,t}(x_{00})\psi_0 \rangle &
\langle \psi_0,j_{0,t}(x_{01})\psi_{\chi_{[0,t]}})\rangle \cr
\langle \psi_{\chi_{[0,t]}}),j_{0,t}(x_{10})\psi_0 \rangle  &
\langle \psi_{\chi_{[0,t]}}),j_{0,t}(x_{11})\psi_{\chi_{[0,t]}})\rangle
\cr}
=\tilde P^t((x_{ij}))$$
its complete positivity follows from Lemma \ref{semi_n}.
The normalization condition
(\ref{norm}) follows from conditions (\ref{4.6}), (47) of Theorem
\ref{th20}.

{\it Sufficiency\/}. Let $\tilde P^t$ be as in the statement of the
theorem and let $(e_{\varepsilon\varepsilon'})$ $(\varepsilon,
\varepsilon'=0,1)$ be a system of matrix units for $M(2,  C)$.
Define for any
$\varepsilon$, $\varepsilon'=0,1$ and $x\in{\cal B}_S$
$$P^t_{\varepsilon,\varepsilon'}(x):=\tilde P^t(x
\otimes e_{\varepsilon,\varepsilon'})$$
Then each $P^t_{\varepsilon,\varepsilon'}$ is a $C_0$--semigroup on
${\cal B}_S$. The complete positivity of $\tilde P^t$ and
condition (\ref{norm}) imply that all the conditions of Theorem \ref{th20}
are satisfied. Therefore the result follows from Theorem \ref{th20}.

\begin{remark}\label{weak_qsde}
The above theorem naturally suggests the following interpretation of the
extended semigroup as matrix elements of a very weak stochastic equation
for the extended flow.
\end{remark}
We may say that a Markov flow
$j_{s,t}:{\cal B}_S\to{\cal B}({\cal H}_S\otimes{\cal F})$
satisfies a {\it weak stochastic equation} in the time interval $[S,T]$ if,
for any sub--interval $[s,t]\subseteq [S,T]$ and for any $\varepsilon$,
$\varepsilon'\in\{0,1\}$, there exist dense subspaces
${\cal D}_{\varepsilon,\varepsilon'}\subseteq{\cal B}_S$
and linear maps
\begin{equation}\label{s5.7}
\kappa_{\varepsilon,\varepsilon'}:{\cal D}_{\varepsilon,\varepsilon'}
\to{\cal B}_S
\end{equation}
such that, for any matrix
$(x_{\varepsilon,\varepsilon'})_{\varepsilon,\var'\in\{0,1\}}
\in M(2,{\cal B}_S)$
with $x_{\var,\var'}\in{\cal D}_{\var,\var'}$, one has
\begin{equation}\label{s5.9}
\pmatrix{
j_{s,t}(x_{00})&j_{s,t}(x_{01})\cr
j_{s,t}(x_{10})&j_{s,t}(x_{11})\cr}=\pmatrix{
x_{00}&x_{01}\cr
x_{10}&x_{12}\cr}+\int_s^t\pmatrix{
j_{s,\tau}(\kappa_{00}(x_{00}))d\tau
&j_{s,\tau}(\kappa_{01}(x_{01}))dA_\tau\cr
j_{s,\tau}(\kappa_{10}(x_{10}))dA^+_\tau
&j_{s,\tau}(\kappa_{11}(x_{11}))dN_\tau\cr}
\end{equation}
where the stochastic integrals in (\ref{s5.9}) are meant componentwise
and the identity means that it holds after taking expectation with
respect to the $M(2,{\cal B}_S)$--valued map:
\begin{equation}\label{s5.8}
\pmatrix{
\langle \psi_0, \ \cdot \ \psi_0 \rangle &
\langle \psi_0,\ \cdot \ \psi_{\chi_{[s,t]}})\rangle \cr
\langle \psi_{\chi_{[s,t]}}),\ \cdot \ \psi_0 \rangle  &
\langle \psi_{\chi_{[s,t]}}),\ \cdot \ \psi_{\chi_{[s,t]}})\rangle  \cr}
\end{equation}
which is completely positive by Lemma \ref{semi_n}.

In this sense we may say that every covariant Markov flow
$j_{s,t}:{\cal B}_S\to{\cal B}({\cal H}_S\otimes{\cal F})$
satisfies a weak stochastic equation and the map
$$
\kappa:=\pmatrix{
\kappa_{00}&\kappa_{01}\cr
\kappa_{10}&\kappa_{11}\cr}
$$
is the generator of the extended semigroup associated to the flow.

Conversely if $\kappa$ is the generator of a completely positive semigroup
in
$M(2,{\cal B}_S)$ then a Markov flow $j_{s,t}:{\cal B}_S
\to{\cal B}({\cal H}_S\otimes{\cal F})$ satisfying equation
(\ref{s5.9}) exists if and only if the element
$\pmatrix{
1&1\cr
1&1\cr}$
is in the domain of $\kappa$ and
$$\kappa\pmatrix{
1&1\cr
1&1\cr}=\pmatrix{
0&0\cr
0&1\cr}$$
%\end{theorem}

\section{Existence of dynamics}

In section 5 we have shown that the theory of (non necessarily homomorphic)
Markov flows can be reduced to the theory of completely positive
semigroups on the algebra $M(2,{\cal B}_S)$ with an additional
normalization condition. In this section we apply this result to the
question of the existence and uniqueness of a homomorphic
Markov flow as solution $j_t$ of a quantum stochastic differential equation
(QSDE) of the form
\begin{equation}\label{QSDE}
dj_t(x)=j_t\circ\sum_\alpha\theta_\alpha(x)dM^\alpha(t)
\end{equation}
with initial condition
\begin{equation}\label{init}
j_0(x)=x
\end{equation}
where $x\in {\cal B}_S$ and the $dM_\alpha(t)$ are the 3 standard boson
Fock stochastic diffferentials,
$$
dM^{-1}(t)=dA(t),\quad dM^{1}(t)=dA^{\dag}(t),\quad dM^{0}(t)=dt,
$$
with the Ito table and conjugation rules
$$
dM^{-1}(t)dM^{1}(t)=dM^{0}(t);\qquad M^{0*}=M^{0},\quad M^{1*}=M^{-1}.
$$
The inclusion of the number process can be dealt with in a similar
technique, but for the application we have in mind (cf. section 7 below)
it is not necessary.
We write the above relations among the stochastic differentials in the
compact notations
\begin{equation}\label{conj}
M^{\alpha *}=M^{\overline\alpha}
\end{equation}
\begin{equation}\label{itotable}
dM^\beta(t)dM^\gamma(t)=\sum_\alpha c^{\beta\gamma}_\alpha
dM^\alpha(t)
\end{equation}
where the $c^{\beta\gamma}_\alpha$ are complex numbers (the structure
constants). Since we are interested in homomorphic flows, we assume that the
maps $\theta_\alpha$ are unital
\begin{equation}\label{kills1}
\theta_\alpha(1)=0\ ;\qquad\forall\,\alpha
\end{equation}
symmetric (the conjugation rules for the $\alpha$'s are the same as for
stochastic differentials)
\begin{equation}\label{howtoconj}
\theta_\alpha(x^*)=\theta_{\overline\alpha}(x)^*\ ;\quad
\forall\,\alpha\ ;\quad\forall\,x\in{\cal B}_S
\end{equation}
and satisfy the stochastic Leibnitz rule:
\begin{equation}\label{stochaderi}
\theta_\alpha(xy)=\theta_\alpha(x)y+x\theta_\alpha(y)+\sum_{\beta,\gamma}
c^{\beta\gamma}_\alpha\theta_\beta(x)\theta_\gamma(y)
\end{equation}
where the structure constants $c^{\beta\gamma}_\alpha$ are the same as
in the Ito table.
\medskip
Our idea is the following: we express the generator of the extended
semigroup $\tilde P^t$ in terms of maps $\theta_\alpha$. Then we take
advantage of the specific properties of these maps to prove that we are
in the conditions to apply the Hille--Yosida theorem giving the
existence and uniqueness of the extended semigroup $\tilde P^t$ and
therefore, because of our results, of the flow itself.

To express the generator of the extended semigroup in terms of the
structure maps let us take the expectation of equation
(\ref{QSDE}) with respect to the functionals $\psi_i$, $\psi_j$, where
$\psi_1=\psi_{\chi[0,T]}$ and the equation is considered in the interval
$[0,T]$. This means that we derive the evolution equations for
$P_{ij}^{'t}=P_{ij}^t$ for $(i,j)=(0,0),(0,1),(1,0)$ and
$P_{11}^{'t}=P_{11}^t e^{T-t}$.

Using the factorization property for exponential vectors
$\psi_f=\psi_{f_{t]}}\otimes\psi_{f_{[t}}$ we get
$$
\langle j_t\circ \sum_\alpha \theta_\alpha(x)dM_\alpha(t) \rangle_{ij}=
\sum_\alpha\langle j_t\circ  \theta_\alpha(x) \rangle_{ij}
\langle dM^\alpha(t)\rangle_{ij},
\quad
\langle\cdot\rangle_{ij}=\langle\psi_{i},\cdot\psi_{j}\rangle
$$
Finally for $P_{ij}^{'t}$ we get
\begin{equation}\label{eq-me}
{d\over dt}\,P_{ij}^{'t}(x)=
P_{ij}^{'t}\circ\sum_\alpha\mu_{ij}^\alpha(t)\theta_\alpha(x),\qquad
\mu_{ij}^\alpha(t)={d\over dt}\langle M^{\alpha} \rangle_{ij}
\end{equation}

The generator $L$ of the semigroup $\tilde P^t$ is given by application, to
the
RHS of the equation (\ref{QSDE}), of the matrix expectation
$(\langle\cdot\rangle_{ij})$ (where $i,j=0,1$
$\langle\cdot\rangle_{ij}=\langle\psi_i,\cdot\psi_j\rangle$, and
$\psi_0$, $\psi_1$ are as above).

Using the known properties of matrix elements of stochastic differentials
between exponential vectors one can express $L$ as a function of the
structure maps $\theta_\alpha$ as follows.
$$
L=\sum_\alpha L^\alpha\theta_\alpha\ ,\quad\alpha=-1,0,1
$$
where
$$
L^0=\pmatrix{
1&1\cr
1&1\cr}\ ,\quad L^{-1}=\pmatrix{
0&1\cr
0&1\cr}\ ,\quad L^{1}=\pmatrix{
0&0\cr
1&1\cr}
$$

More explicitly:
\begin{equation}\label{L}
L=\pmatrix{
\theta_0,&\theta_0+\theta_{-1}\cr
\theta_0+\theta_{1},&\theta_0+\theta_{1}+\theta_{-1}\cr}
\end{equation}
where the matrix $L$ acts elementwise on $M(2,{\cal B}_S)$:
$ (L_{ij})(x_{ij})=( \ L_{ij}(x_{ij}) \ )$.
For the calculations in the following Lemma it is convenient to
introduce the following notations: for the action of $L^\alpha$ on a
generic $2\times2$--matrix $x=(x_{ij}) $ we use the convenction
$$L^0(x)=x \ \quad L^1(x)=\pmatrix{
0&0\cr
0&1\cr}x\ ,\quad L^{-1}(x)=x\pmatrix{
0&0\cr
0&1\cr}
$$
where the products on the right hand sides are interpreted as usual
matrix multiplications.

The following lemma is important to reduce the problem of
existence, uniqueness and conservativity of a flow to an
application of the Hille--Yosida theorem.

\medskip

\begin{lemma}\label{lemma22}
If the structure maps $\theta_\alpha$ in (\ref{QSDE}) have a common core
${\cal B}_0$ which is a $*$--subalgebra invariant under the square root,
then the operator $L$ is symmetric and
generates a completely positive semigroup.

%exists a real constant $c$ such
%that $S=L-c$ is completely dissipative.

\end{lemma}

\noindent{\it Proof\/}.\qquad
The symmetry is obvious because the complete positivity of the semigroup
$\tilde P^t$ implies that: $\tilde P^t(x^*)=\tilde P^t(x)^*$.

To prove the existence of the semigroup with the generator $L$
we will use the Hille--Yosida theorem (cf. \cite{BR}).
To use this theorem we have to prove the dissipativity of the generator.

To prove this we
will use that $\theta_{\pm 1}$ are mutually adjoint derivations and
$\theta_0$ is a dissipation satisfying the property
$$
\theta_0(AB)=\theta_0(A)B+A\theta_0(B)+\theta_{-1}(A)\theta_1(B)
$$
that is exactly formula (\ref{stochaderi}).
Applying the generator $L$ to $2\times 2$--matrix with the entries in
${\cal B}_S$ we get
$$
L(x^*x)=(\theta_{-1}(x^*)x+x^*\theta_{-1}(x))\pmatrix{ 0&0\cr
0&1\cr}+
$$
$$
+\pmatrix{
0&0\cr
0&1\cr}(\theta_1(x^*)x+x^*\theta_1(x))+$$
$$+\theta_0(x^*)x+x^*\theta_0(x)+\theta_{-1}(x^*)\theta_1(x)=$$
$$=L(x^*)x+x^*L(x)+\theta_{-1}(x^*)\theta_1(x)+$$
$$+\theta_{-1}(x^*)\left[x,\pmatrix{
0&0\cr
0&1\cr}\right]+\left[\pmatrix{
0&0\cr
0&1\cr}, x^*\right]\theta_1(x)=
$$
\begin{equation}\label{for_dissip}
=L(x^*)x+x^*L(x)+\left|\theta_1(x)+\left[x,\pmatrix{
0&0\cr
0&1\cr}\right]\right|^2-\left|\left[x,\pmatrix{
0&0\cr
0&1\cr}\right]\right|^2
\end{equation}
where we use the notation $|A|^2=A^*A$. The third term in (\ref{for_dissip})
is nonnegative, therefore
\begin{equation}\label{c4}
L(x^*x)\geq L(x^*)x+x^*L(x)-|\delta(x)|^2
\end{equation}
with $\delta$ given by
$$
\delta(x)=i\left[x,\pmatrix{
0&0\cr
0&1\cr}\right]
$$

From the identity
$$
\delta^2(x^*x)=\delta(\delta(x^*)x+x^*\delta(x))=\delta^2
(x^*)x+2\delta(x^*)\delta(x)+x^*\delta^2(x)
$$
and from the fact that $\delta$ is a symmetric
derivation, it follows that
\begin{equation}\label{c5}
|\delta(x)|^2={1\over2}\,(\delta^2(x^*x)-\delta^2(x^*)x-x^*\delta^2
(x))
\end{equation}
Given (\ref{c5}), (\ref{c4}) is equivalent to
\begin{equation}\label{c6}
\left(L+{1\over2}\,\delta^2\right)(x^*x)\geq\left(L+{1\over2}\,
\delta^2\right)(x^*)x+x^*\left(L+{1\over2}\,\delta^2\right)
(x)
\end{equation}

%Now one finds
%$$\delta\pmatrix{
%x_{11}&x_{12}\cr
%x_{21}&x_{22}\cr}=i\pmatrix{
%0&x_{12}\cr
%-x_{21}&0\cr}$$
%Therefore
%$$\delta^2\pmatrix{
%x_{11}&x_{12}\cr
%x_{21}&x_{22}\cr}=-\pmatrix{
%0&x_{12}\cr
%x_{21}&0\cr}$$
%keeping in mind the form (\ref{L}) of $L$ we get
%$$
%\left(L+{1\over2}\,\delta^2\right)\pmatrix{
%z_{11}&z_{12}\cr
%z_{21}&z_{22}\cr}=\pmatrix{
%\theta_0(z_{11})&(\theta_0+\theta_{-1})(z_{12})-{1\over2}\,z_{12}\cr
%(\theta_0+\theta_{+1})(z_{21})-{1\over2}\,z_{21}&
%(\theta_0+\theta_{+1}+\theta_{-1})(z_{22})\cr}
%$$

Because all maps $\theta^\alpha$ have a common core ${\cal B}_0$, that is a
$*$--subalgebra, invariant under the square root,
according to (3.2.22) of \cite{BR}
it follows that $L+{1\over2}\,\delta^2$ is a closable dissipation.
By similar arguments one can prove complete dissipativity.

By the Hille--Yosida theorem (cf. \cite{BR})
the dissipative operator
$$
S=L+{1\over2}\,\delta^2=\pmatrix{
\theta_0&\theta_0+\theta_{-1}-{1\over2}\cr
\theta_0+\theta_1-{1\over2}&\theta_0+\theta_{+1}+\theta_{-1}\cr}
$$
generates a semigroup of contractions $P^t$ on $M(2,{\cal
B}_S)$.

The proof of this theorem is by
the resolvent approximation.  In this approximation the semigroup $e^{tS}$
is
the strong limit of the semigroups $e^{tS_{\varepsilon}}$  with the bounded
generator
$$
S_{\varepsilon}=S(1-\varepsilon S)^{-1}=
-\varepsilon^{-1}(1-(1-\varepsilon S)^{-1})
$$
But by \cite{Lindblad} a bounded symmetric completely dissipative operator
on a $W^*$--algebra generates a completely positive (contractive) semigroup.
Since $e^{tS_{\varepsilon}}$ converges strongly to $e^{tS}$, also
the semigroup $e^{tS}$ is completely positive.

The operators $L$ and ${1\over2}\,\delta^2$ commute.
Now ${1\over2}\,\delta^2$ is the generator of the $C^0$--semigroup
$$
e^{{t\over2}\,\delta^2}\pmatrix{
x_{11}&x_{12}\cr
x_{21}&x_{22}\cr}=\pmatrix{
x_{11}&e^{-{t\over2}}x_{12}\cr
e^{-{t\over2}}x_{21}&x_{22}\cr}
$$

We get that
$$
e^{t\left(L+{1\over2}\,\delta^2\right)}e^{-{t\over2}\,\delta^2}
$$
is a completely positive semigroup with generator $L$.

This finishes the proof of the Lemma.

\medskip

\begin{lemma}\label{lemma23}
The semigroup $\tilde P^t=e^{tL}$ satisfies
$$
e^{tL}\pmatrix{1&1\cr1&1\cr}=
\pmatrix{1&1\cr1&1\cr};\qquad
\left(e^{tL}\right)'\pmatrix{1&1\cr1&1\cr}=
\pmatrix{1&1\cr1&e^t\cr}
$$
\end{lemma}

\noindent{\it Proof\/}.\qquad

The semigroup $e^{tL_{\varepsilon}}$ is
given by the iterated series
$$
e^{tL_{\varepsilon}}=\sum_{k=0}^{\infty}
{t^k\over k!}L_{\varepsilon}^k
$$
where the regularization $L_{\varepsilon}$ have the form
$$
L_{\varepsilon}=L(1-\varepsilon L)^{-1}=
-\varepsilon^{-1}(1-(1-\varepsilon L)^{-1})
$$
From formula (\ref{L}) it follows that the generator $L$ and therefore
its regularization $L_{\varepsilon}$ kills $\pmatrix{1&1\cr1&1\cr}$
(because each of the $\theta_{\alpha}$ kills $1$). Therefore from the
form of the iterated series it follows that the semigroup
$e^{tL_{\varepsilon}}$ is conservative ($e^{tL_{\varepsilon}}(1)=1$).
Applying to $e^{tL_{\varepsilon}}\pmatrix{1&1\cr1&1\cr}$
the limit $\varepsilon\to 0$ and using that
$e^{tL}$ is the strong limit of $e^{tL_{\varepsilon}}$ we get the
conservativity of $e^{tL}$. Using the rescaling
$P_{11}^{'t}=P_{11}^t e^{T-t}$ we get the statement of the lemma.

\bigskip

\begin{theorem}\label{th24}
In the assumptions of Lemma \ref{lemma22}, there exists a unique flow
$j_t(x)$
whose extended semigroup is $e^{tL}$. Moreover $j_t(x)$ is the unique
solution of the Cauchy problem
(\ref{QSDE}), (\ref{init}).
\end{theorem}

\noindent{\it Proof\/}.\qquad
Because of Lemma \ref{lemma22}, $L$ is the generator of a semigroup $e^{tL}$
satisfying the conditions of Theorem \ref{ext_semi}.
Hence there exists a unique flow $j_t(x)$, whose extended semigroup
is $e^{tL}$.
Therefore, if we prove that this flow
satisfies equation (\ref{QSDE}) with initial condition (\ref{init})
then, by the uniqueness of the matrix elements in the exponential vectors,
the
uniqueness of the solution follows.

First we prove that the flow $j_t(x)$ satisfies (\ref{QSDE}), (\ref{init}).
To this goal it is sufficient to prove that
for a dense set of vectors $\psi$, $\psi'$ the matrix element
$\langle\psi, j_t(x)\psi'\rangle$ satisfies the equation obtained by
taking $\langle\psi, \cdot\psi'\rangle$--expectation of equation
(\ref{QSDE})
and moreover $\langle\psi, j_t(1)\psi'\rangle=1$.
But this is true because of (\ref{eq-me}).
This finishes the proof of the theorem.

\section{Application to quantum Glauber dynamics}

In this section we apply our technique to prove the existence of the flow
for a quantum system of spins on a lattice.
Starting with the work of Glauber \cite{Glauber}
the dynamics of infinite classical lattice systems has been considered
by many authors and has led to study the
ergodic and equlibrium properties of a new class of
classical Markov semigroups (cf. \cite{Liggett} for a general survey and for
further references). Quantum analogues of these semigroups have also
been considered by several authors (e.g. \cite{Martin}, \cite{Matsui},
\cite{MajZeg96a}, \cite{Spitzer}, \cite{Sullivan}, \dots).
However the problem of deriving these Markovian
semigroups, and more generally the stochastic flows, as limits of
Hamiltonian systems, was open both in the classical and in the quantum case.
This problem was solved in the paper \cite{AcKo99a} where a
quantum stochastic differential equation (QSDE)
describing the dynamics of a infinite volume spin system interacting with
bosonic white noise was derived.
In the simplest case ($1$--dimensional lattice with nearest neighbor,
translation invariant interaction) we have the following picture:
the spins (or more generally two state systems) are enumerated
by integer numbers and
the dynamics of the system (the flow) is given by the following
QSDE (Langevin equation)
\begin{equation}\label{Langevinforspins}
dj_t(X)=\sum_{\alpha=-1,0,1}j_t\circ \theta_{\alpha}(X) dM^{\alpha}(t)
\end{equation}
where
$$
dM^{-1}(t)=dB(t),\quad dM^{1}(t)=dB^{*}(t)
$$
$$
\theta_{-1}(X)=-i[X,F^{(++)*}_{\Lambda}], \quad
\theta_{1}(X) =-i [X,F^{(++)}_{\Lambda}]
$$
$$
\theta_0(X)=
\left(
\theta_{0}^{(0,-1)}+
\theta_{0}^{(0,1)}+
\theta_{0}^{(-1)}+
\theta_{0}^{(1)}
\right)(X)=
$$
$$
=\biggl(
\sum_{\varepsilon,\mu}\left(
-i \hbox{Im} \,(g|g)^-_{(\varepsilon\mu)}
[X, F_{\Lambda}^{\varepsilon\mu*} F_{\Lambda}^{\varepsilon\mu} ]
+i \hbox{Im} \,(g|g)^+_{(\varepsilon\mu)}
[X, F_{\Lambda}^{\varepsilon\mu} F_{\Lambda}^{\varepsilon\mu *}]
\right)+
$$
$$
+\hbox{Re}\,(g|g)^-_{(++)}
\left(
2F^{(++)*}_{\Lambda}X F^{(++)}_{\Lambda}-
\{X,F^{(++)*}_{\Lambda}F^{(++)}_{\Lambda}\}\right)+
$$
$$
+\hbox{Re}\,(g|g)^+_{(++)}
\left(
2F^{(++)}_{\Lambda}XF^{(++)*}_{\Lambda}-
\{X,F^{(++)}_{\Lambda}F^{(++)*}_{\Lambda}\}
\right)
$$
where $(g|g)^{\pm}_{(\varepsilon\mu)}$ are constants with
non--negative real part (whose explicit form is given in \cite{AcKo99a})
and the stochastic differentials
satisfy the following Ito table
$$
dB(t)dB^*(t)=\hbox{Re}\,(g|g)^+_{(++)} dt
$$
$$
dB^*(t)dB(t)=\hbox{Re}\,(g|g)^-_{(++)} dt
$$
The operators $F^{(\varepsilon,\mu)}_{\Lambda}$
acting on the spin degrees of freedom have the form
$$
F^{(\varepsilon\mu)}_{\Lambda}=
\sum_{r\in\Lambda}F^{(\varepsilon,\mu)}_{r}
$$
where $\Lambda$ is a subset of the lattice of integer numbers and
$$
F^{(++)}_r=|1_{r-1}\rangle\langle 1_{r-1}|
|1_{r}\rangle\langle -1_{r}||1_{r+1}\rangle\langle 1_{r+1}|+
|-1_{r-1}\rangle\langle -1_{r-1}|
|-1_{r}\rangle\langle 1_{r}||-1_{r+1}\rangle\langle -1_{r+1}|
$$
and the other operators $F^{(\varepsilon,\mu)}_r$ are defined
correspondingly. For fixed spin number $r$ the operators
$F^{(\varepsilon\mu)}_r$ are indexed by
configurations of nearest neighbors of the spin at $r$
(for every $r$ we have 4 configurations).
We denote these configurations $++$, $+-$, $-+$ and $--$
(the first symbol is the orientation of the spin on the left
of $r$ and the second --- on the right).

One can prove that the structure maps $\theta_{\alpha}$
in (\ref{Langevinforspins}) satisfy the conditions of lemma
\ref{lemma22} and conditions (\ref{stochaderi}), (\ref{kills1}),
(\ref{howtoconj}). Therefore we can apply to the stochastic
dynamics of the considered system of spins the approach developed in
the present paper and prove the existence of the flow which solves the
QSDE (\ref{Langevinforspins}). These considerations
continue to hold also for multidimensional lattices, but in this case
the Langevin equation is more complex \cite{AcKo99a}.

\bigskip

\noindent
{\bf Acknowledgements.}

S.Kozyrev is grateful to Luigi Accardi and Centro Vito Volterra
where this work was done for kind hospitality.
This work was partially supported by INTAS 96-0698 grant.
S.Kozyrev is also supported by RFFI 990100866 grant.

\end{document}